\documentclass[acmsmall]{acmart}

\usepackage{adjustbox}
\usepackage{xr}
\usepackage{hyperref}

\usepackage[normalem]{ulem}
\usepackage{amsmath}

\newcommand{\amanat}[1]{} 
\newcommand{\update}[1]{\textcolor{black}{#1}}

\newcommand{\dhruv}[1]{} 

\newcommand{\co}[1]{}
\newcommand{\remove}[1]{}

\usepackage{listings}
\usepackage{xcolor}

\AtBeginDocument{%
  \providecommand\BibTeX{{%
    \normalfont B\kern-0.5em{\scshape i\kern-0.25em b}\kern-0.8em\TeX}}}

\setcopyright{acmlicensed}
\acmDOI{10.1145/3698135}
\acmYear{2024}
\acmJournal{PACMHCI}
\acmVolume{8}
\acmNumber{ISS}
\acmArticle{535}
\acmMonth{12}
\received{2024-02-22}
\received[accepted]{2024-05-30}

\begin{document}

\title[Exploring Pointer Enhancement Techniques for Target Selection on Large Curved Display]{{Exploring Pointer Enhancement Techniques for Target Selection on Large Curved Display}}

\author{Dhruv Bihani}
\affiliation{%
  \institution{University of British Columbia, Okanagan}
  \city{Okanagan, BC}
  \country{Canada}}
\email{dhruv109@student.ubc.ca}
\orcid{0009-0006-6498-766X}
\authornote{Both authors contributed equally to this research.}

\author{A. K. M. Amanat Ullah}
\affiliation{%
  \institution{University of British Columbia, Okanagan}
  \city{}
  \country{Canada}}
\email{amanat7@student.ubc.ca}
\orcid{0000-0001-5402-0160}
\authornotemark[1]

\author{Charles-Olivier Dufresne-Camaro}
\affiliation{%
  \institution{University of British Columbia, Okanagan}
  \city{Kelowna, BC}
  \country{Canada}}
\email{chcamaro@student.ubc.ca}
\orcid{0000-0002-8563-7523}

\author{William Delamare}
\affiliation{%
  \institution{Univ. Bordeaux, ESTIA-Institute of Technology, Bidart}
  \city{}
  \country{France}}
\email{william.delamare@acm.org}
\orcid{0000-0002-1830-4294}

\author{Pourang Irani}
\affiliation{%
  \institution{University of British Columbia, Okanagan}
  \city{}
  \country{Canada}}
\email{pourang.irani@ubc.ca}
\orcid{0000-0002-7716-9280}

\author{Khalad Hasan}
\affiliation{%
  \institution{University of British Columbia, Okanagan}
  \city{}
  \country{Canada}}
\email{khalad.hasan@ubc.ca}
\orcid{0000-0002-4815-5461}

\renewcommand{\shortauthors}{Bihani, Ullah, Dufresne-Camaro, Delamare, Irani, and Hasan }

\begin{abstract}
  
Large curved displays are becoming increasingly popular due to their ability to provide users with a wider field of view and a more immersive experience compared to flat displays. Current interaction techniques for large curved displays often assume a user is positioned at the display's centre, crucially failing to accommodate general use conditions where the user may move during use. In this work, we investigated how user position impacts pointing interaction on large curved displays and evaluated cursor enhancement techniques to provide faster and more accurate performance across positions. To this effect, we conducted two user studies. First, we evaluated the effects of user position on pointing performance on a large semi-circular display (3m-tall, 3270R curvature) through a 2D Fitts' Law selection task. Our results indicate that as users move away from the display, their pointing speed significantly increases (at least by 9\%), but accuracy decreases (by at least 6\%). Additionally, we observed participants were slower when pointing from laterally offset positions. Secondly, we explored which pointing techniques providing motor- and visual-space enhancements best afford effective pointing performance across user positions. Across a total of six techniques tested, we found that a combination of acceleration and distance-based adjustments with cursor enlargement significantly improves target selection speed and accuracy across different user positions. Results further show techniques with visual-space enhancements (e.g., cursor enlargement) are significantly faster and more accurate than their non-visually-enhanced counterparts. Based on our results we provide design recommendations for implementing cursor enhancement techniques for large curved displays.

\end{abstract}



\begin{teaserfigure}
  \includegraphics[width=\textwidth]{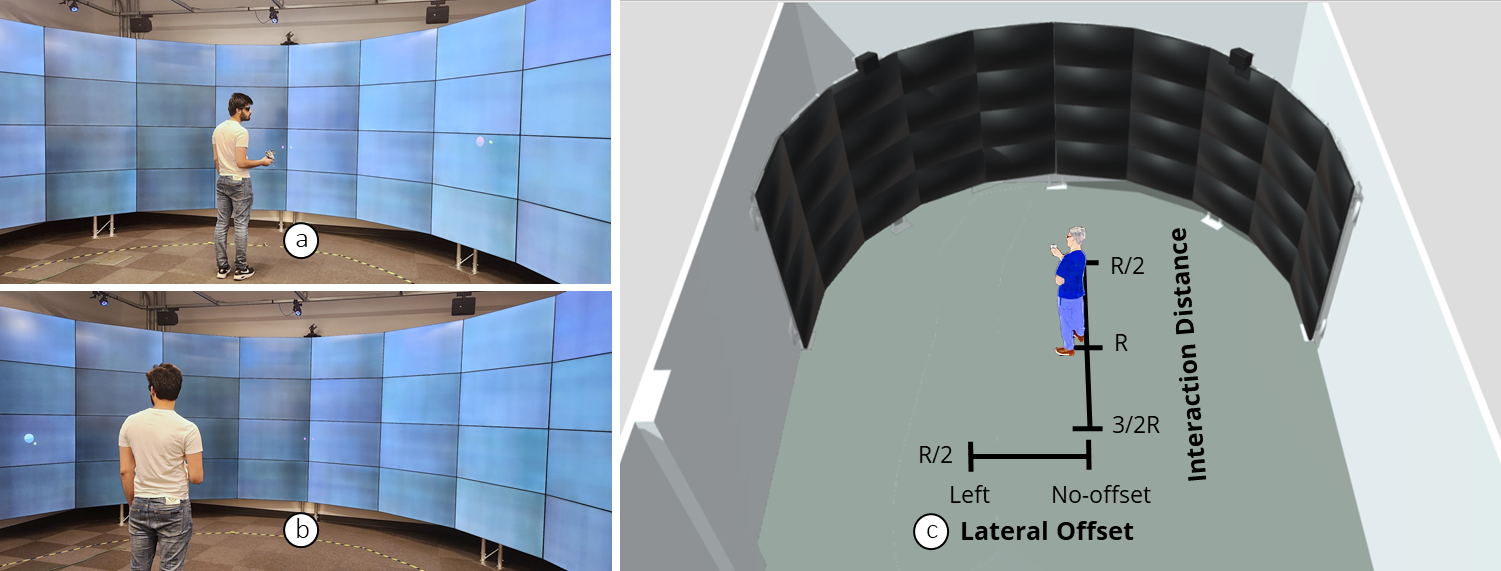}
\caption{(a-b) A user selects targets on an immersive 3D large curved display. (c) \textbf{Interaction Distance and Lateral Offset}: Interaction Distance represents user positions along the main axis, while Lateral Offset represents users positions with a deviation of R/2 towards the left from the main axis.}
  \Description{(a-b) A user selects targets on an immersive 3D large curved display. (c) The user is positioned at different distances from the centre of the display. }
  \label{fig:teaser}
\end{teaserfigure}

\maketitle 


\section{Introduction}

Large displays, whether flat or curved, are becoming increasingly popular in fields like data visualization \cite{novak2008designing, satyanarayan2013chi, birnholtz2007exploratory}, entertainment \cite{ardito2015interaction, scheible2005mobilenin, khoo2009designing}, and collaborative applications \cite{andrews2011information, card1999information, baudisch2001focus, cockburn2009review}. 
Curved displays have been found to enhance user satisfaction and task performance compared to large flat displays, because they offer a wider field of view \cite{ahn2014research, kyung2021curved, sultana2024exploring}.
Pointing using ray-casting is a common interaction technique on large displays because it allows for rapid cursor movements and for users to move freely in front of the screen \cite{kopper2008increasing}. 
User pointing performance on large flat displays has been shown to vary in speed and accuracy based on notably the user's \textit{Interaction Distance}, i.e., the distance from the user perpendicular to the centre of the display \cite{kopper2010human,janzen2016modeling,kovacs2008perceptual,tao2021freehand, sindhupathiraja2024exploring}.
The size of the selection target, as well as the distance to the target, further affects performance, with small distant targets leading to slower selection times and lower accuracy \cite{tao2021freehand, babic2018pocket6, siddhpuria2018pointing, nancel2013high}. 

Pointing on curved displays introduces unique challenges due to the geometry of the displays altering how users may interact with them as they move around (e.g., laterally from the display's centre) \cite{zannoli2017perceptual}.
For instance, interacting with a curved display often requires more head rotation to engage with different areas, compared to flat displays \cite{shupp2009shaping}.
Users' field-of-view during interactions also varies with user position when the display is curved \cite{zannoli2017perceptual}. 
Furthermore, curved displays cannot be represented as 2D planes, which limits the applicability of commonly-used pointer enhancement techniques \cite{casiez2008impact,nancel2013high} which rely on planar geometry to adjust the effects of motor movement. 


\update{Handheld controller-based ray-casting is a widely used technique for pointing on large displays, where users manipulate a ray in 3D space by moving the controller with their hand. Although other research leverages head and eye movements for pointing at large displays \cite{khamis2017eyescout, James_2020_personal}, hand-based pointing remains the most prevalent technique due to its intuitive nature and ease of learning, which lead to faster and more precise selections \cite{lin2015investigation, bernardos2016comparison}. Nevertheless, hand-based pointing is susceptible to precision errors due to hand tremors and involuntary movements \cite{myers2002interacting}. 
} 
User position has additionally been shown to significantly affect pointing performance on flat displays \cite{kopper2010human,janzen2016modeling,kovacs2008perceptual,tao2021freehand, hourcade2012small, Ullah2023Pointing}, resulting in less accurate selections at larger pointing distances.
Numerous techniques have been developed over the years to improve pointing performance by modifying the ratio between cursor movement and hand movement \cite{casiez2008impact, nancel2013high, chowdhury2022wriarm, chowdhury2023paws}, and modifying the visuals \cite{GrossmanCHI2005}.
However, we find no prior work on designing techniques that leverage \textit{Interaction Distance} for large displays, whether flat or curved. 
More importantly, the performance of commonly-used cursor enhancement techniques such as cursor acceleration \cite{casiez2008impact,nancel2013high} and enlargement \cite{GrossmanCHI2005} have not been investigated for curved displays. 
Considering the unique design space of large curved displays, existing techniques may not generalize well as-is to this space.

In this work, we aim to understand 1) how user position affects pointing performance on large curved displays, and 2) which cursor enhancement techniques best afford users with optimal pointing performance across positions by ensuring faster selections and higher accuracy. 
To this effect, we conduct two user studies.
First, we evaluate how \textit{Interaction Distance}, and \textit{Lateral Offset} (i.e., lateral distance from the display's centre) affect pointing performance on a large curved display through a Fitts' Law selection task.
We confirm the effect of \textit{Interaction Distance} is similar to large flat displays: larger distances increase pointing speed but lower pointing accuracy \cite{kopper2010human}. 
Furthermore, we observed a significant effect of \textit{Lateral Offset}  on pointing speed, where selections are slower from offset positions. 
Secondly, we explore six cursor enhancement techniques defined along two axes --- motor space (i.e., motor-space acceleration and distance-based augmentations) and visual space (i.e., cursor enlargement) --- to identify which technique best reduces performance imbalances across different user positions (i.e., \textit{Interaction Distances} and \textit{Lateral Offsets}). 
We found that \textit{Interaction Distance} combined with controller-acceleration-based techniques led to superior pointing performance. 
Furthermore, augmenting motor-space enhancements with visual-space enhancements improves pointing accuracy and speed.

The main contributions of our work are: 
\begin{itemize}
    \item Empirical support for the effects of user position (i.e., \textit{Interaction Distance} and \textit{Lateral Offset}) on pointing performance on large curved displays. 
    \item Evidence of the benefits of combining acceleration-based, and \textit{Interaction-Distance}-based cursor enhancement techniques on pointing performance on large curved displays across user positions.
    \item Design recommendations on improving pointing performance on large curved displays using motor-space and visual-space enhancements. 
    
\end{itemize}

\section{Background and Related Work}



\subsection{Large Displays: Flat vs Curved}


Comparative studies between flat and curved large displays have been conducted across various tasks such as visual searches \cite{kyung2021curved, shupp2009shaping, sultana2024exploring}, data visualization \cite{liu2020design, shupp2009shaping} and reading \cite{wei2020reading, park2019effects, sultana2024exploring}.
Curved displays have been shown to be more effective on graph comparison and map search tasks \cite{shupp2009shaping} and video watching \cite{urakami2021comparing, choi201553, ahn2014research}, notably due to their improved ergonomics.  
Zannoli et al. \cite{zannoli2017perceptual} have additionally been found to afford a wider range of user positions for interactions and reduce reflections, but require more complex rendering techniques to support their geometry \update{for a 65-inch with large display curvature of 4.18m}. 
Interacting with curved displays also require more head movements, and are less suited for overviews \cite{shupp2009shaping} \update{for large curved (having 0.76m radius) and flat display with the combination of 24
monitors arranged in 8 × 3 grid approximately 9 feet wide and 3 feet tall}. 

Display types have also been compared based on content density (e.g., the number of charts displayed simultaneously) \cite{liu2020design, shupp2009shaping}.
Users performed better on flat displays for small numbers of data visualizations rendered on a 2D grid \cite{liu2020design}, but preferred curved displays when content density increased. 
Furthermore, Kuyung and Park \cite{kyung2021curved} found that large semi-circular \update{displays (33" and 50") with a curvature radius of 60cm} lead to improved performance in visual search tasks, compared to flat and other curved displays.
Additionally, reading on \update{27" displays} was found to be faster when the users were positioned at a distance equal to the curved display curvature's radius \cite{park2019effects}.
VR studies further revealed horizontally curved virtual displays to be more comfortable than vertically curved, flat and spherical displays \cite{wei2020reading}.



Overall, curved displays offer immersive viewing experiences and can enhance task performance.
However, they introduce unique and open challenges compared to flat displays in terms of supporting user interactions across various positions, which have yet to be solved.
Therefore, in this work, we investigate how the performance of pointing --- a general task --- varies on large curved displays, and what sets of techniques can best improve user performance in this context.

\subsection{Pointing on \update{Large Curved} Displays } 

Pointing is considered one of the most common metaphors for selection and interaction with user interfaces \cite{bowman20043d, argelaguet2009efficient}.
Ray-casting (i.e., a ray projected from the tip of an input device) is the most widely used method for pointing at targets on large \update{flat} displays \cite{cavens2002interacting, davis2002lumipoint, oh2002laser, jota2010comparisonGIpaper}.
Research on pointing on large displays has primarily focused on large flat displays \cite{sluyters2023consistent, jota2010comparisonGIpaper, nancel2013high, siddhpuria2018pointing, haque2015myopoint}. 
We thus include in this section relevant work on pointing at \remove{non-large} \update{large curved} displays.

We found limited prior work on pointing at \update{large} curved displays \cite{Ullah2023Pointing, Ullah2023LargeCurved, hennecke2013investigating, ens2016moving}.
Pointing performance has been shown to not vary between physical and virtual \update{large} curved displays \cite{Ullah2023Pointing}, facilitating the exploration of this design space.
Fitts' Law was found to be applicable for 2D pointing tasks on \update{large} curved displays using direct touch or mouse \cite{hennecke2013investigating}.
Similarly, 1D pointing performance has been shown to vary on \update{large} virtual curved displays based on curvature and improve when the display is less curved \cite{Ullah2023LargeCurved}.
Additionally, the speed-accuracy trade-off in pointing has been shown to apply to 1D pointing on \update{large} curved displays, but lateral displacements were not found to affect performance \cite{Ullah2023Pointing}.
\remove{Finally, larger field-of-views (e.g., having the user stand closer to the curved display) may improve pointing performance \cite{ens2016moving} for CAVE environment but require further investigation regarding user position and display properties.} 

Taken together, our understanding of the effects of factors unique to curved displays (e.g., user position relative to display curvature) on pointing performance is limited.
In this work, we first validate the effects of \textit{Interaction Distance} and \textit{Lateral Offset} on 2D pointing task performance on large  curved displays, and then investigate cursor enhancement techniques to best support pointing across varied user positions.

\subsection{Cursor Enhancement}

We explore in this work which cursor enhancement techniques best improve the accuracy and selection times for pointing on a large curved display from various positions.
To the best of our knowledge, no prior work has explored techniques for this setting.
We thus review existing literature on cursor enhancement techniques on large flat displays to identify candidate techniques. 

We categorize cursor enhancement techniques under two groups: \textbf{motor-space} enhancements, and \textbf{visual-space} enhancements. 
\textbf{Visual-space} enhancements act on cursor and target sizes to facilitate target selection.
For example, \textit{bubble cursor} which dynamically updates its size based on the proximity of surrounding targets \cite{GrossmanCHI2005}. 
\textbf{Motor-space} enhancements instead act on the cursor's speed to improve pointing performance. 
Such techniques often modify the Control-to-Display (CD) gain to adjust the ratio between cursor movement and input device movement to require smaller input movements to traverse large distances, or larger input movements when the cursor is near a target to facilitate accurate selections \cite{casiez2011no}.
Common motor enhancement techniques involve Constant Gain (CG), Pointer Acceleration (PA) which dynamically adjusts CD gain based on input device movement (e.g., a mouse) \cite{casiez2008impact}, and Adaptive Pointing (AP) which adjusts the CD gain based on input device speed and device-cursor offset \cite{konig2009adaptive}.
PA has notably been shown to lead to faster selections when their CD gain levels are not precisely optimized \cite{casiez2008impact}.
AP further leads to more accurate and faster selections than direct pointing techniques \cite{konig2009adaptive}.
Techniques combining ray-casting (absolute pointing) with relative pointing have also been proposed \cite{VogelUIST2005,McCallumUIST2009,ForlinesUIST2006,GalloIEEE2010}. 
Finally, cursor enhancements can be designed for high-precision settings, such as RelaSmall and RelaLarge which adjust CD gain based on input device acceleration \cite{nancel2013high}.

To the best of our knowledge, no prior work has investigated cursor enhancement on large curved displays. 
We argue relying on performance and guidelines for large flat displays to be insufficient due to the unique design space defined by large curved displays: for instance, changes in field-of-views as users move away from the display's centre, and non-planar geometry.
We additionally note that although user position has been found to impact interaction on curved displays \cite{zannoli2017perceptual, Ullah2023Pointing}, no cursor enhancement techniques for large displays have been designed and evaluated considering user position.
 In this work, we conduct an investigation of six techniques to support pointing on large curved displays inspired by existing motor-space and visual-space techniques. 

to\section{User Study 1 - Pointing Performance across User Positions} 

Pointing performance on flat displays has been shown to vary based on the user's position with respect to the display \cite{kopper2010human,janzen2016modeling,kovacs2008perceptual,tao2021freehand, hourcade2012small}.
The effects on large curved displays have also been studied, and only the the distance of the user from the display, and not any lateral displacements was shown to impact performance \cite{Ullah2023Pointing}.
However, the task (1D pointing with reciprocal targets at fixed amplitudes) may have been too simple, and limited the impacts of user position on performance.
In order to design cursor enhancement techniques that best support users on large curved displays, we see a need to first verify the effects of user position on pointing performance.


We therefore conducted a within-subject user study to evaluate how the user's position, defined here as a 2D grid around the display (see Fig. \ref{fig:teaser}c), affects their pointing performance on a large curved display.
Specifically, we employed a 2D Fitts' Law selection task to investigate the effects of two factors on pointing performance: 1) the user's \textit{Interaction Distance} (i.e., their perpendicular distance to the centre of the large curved display), and 2) the user's \textit{Lateral Offset} from the centre of the display.

Formally, we formulate the following Hypotheses around the two position-based factors:
\begin{itemize}
    \item \textbf{H1}) Pointing speed will increase as the \textit{Interaction Distance} increases, as users will require smaller movements to cover large distances.
    \item \textbf{H2}) Pointing performance (speed and accuracy) will be optimal when the user is aligned with the display's centre, compared to when their position includes a \textit{Lateral Offset}. 
   \item \textbf{H3)} Pointing accuracy will decrease as \textit{Interaction Distance} increases, as moving farther away amplifies hand jitter \cite{kopper2011understanding}. 
\end{itemize}


\subsection{Factors}

This study focuses on the effects of two factors: \textit{Interaction Distance} and \textit{Lateral Offset} (see Fig. \ref{fig:teaser}c).
\textit{Interaction Distance} represents how far the user is in front of the display when considering the display's \textit{main axis}, i.e., the axis passing through the curvature center and the middle of the display. 
\textit{Lateral Offset} corresponds to the user's lateral displacement (left or right) relative to the main axis.
Furthermore, as we use a 2D Fitts' Law selection task, we also consider the effects of the selection target's \textit{Amplitude} and \textit{Width} on pointing performance.

\textbf{Interaction Distance}: We consider three distances relative to the radius of the display's curvature ($R = 3.27m$): \(\frac{1}{2}\)R (1.64m), R (3.27m), and \(\frac{3}{2}\)R (4.91m).
Using multiples of $R$ allows our results to generalize beyond the specific large curved display used here.

\textbf{Lateral Offset}: Since target position on axis-symmetric directions has no difference in pointing performance \cite{zhang2012extending,radwin1990method}, we consider two offsets: No-offset where a participant is positioned along the \textit{main axis}, and Left-offset where the participant is positioned \(\frac{1}{2}\)R (1.64m) to the left of the \textit{main axis}.
We further limit recruitment to strictly right-handed participants to control for potential differences due to handedness and studying only left offsets.
User position is thus examined through six ($2\times3$) distinct positions in front of the display (see Fig. \ref{fig:teaser}c). 

\textbf{Target Amplitudes $\And$ Width}: We evaluate pointing performance through three target amplitudes (i.e. distance between the targets): 250cm, 500cm, and 750cm.
Additionally, we consider two target widths (i.e., the diameter of the spherical targets): 20cm, and 70cm. 
This results in six distinct Indexes of Difficulty (IDs) ranging from 2.85 to 6.25 bits.  
The specific levels were selected based on the arc length of the large curved display along its surface, and fine-tuned through a pilot study.
These conditions mimic real-world situations where interface elements might appear anywhere on the curved screen, independently of the user's current position.

\subsection{Apparatus}

We conducted our study on a large physical curved display produced by Mechdyne \cite{mechdyne-new1}. 
The display is semi-circular ($180^\circ$ viewing angle), 3m tall, and has a radius of 3.27m.  
\update{Here 3270R refers to a screen having a radius of 3270 millimeters or 3.27 meters.} 
The display consists of forty 46-inch LED 3D displays set in a 4x10 grid (see Fig. \ref{fig:teaser}a-c). 

We implemented the pointing task using the Unity game engine \cite{unity}. 
A ray, corresponding to the user's current pointing direction, was rendered from a Quest 2 controller held by the user to point at the targets (see Fig. \ref{fig:teaser}a-b). 
The ray extended from the controller in the orthogonal direction and a cursor was placed at the endpoint of the ray on the large curved display (Figure \ref{fig:teaser}a-b).
Target selection was done by pressing the controller's trigger button.
Markers were attached to the controller to track its position via an 8-camera Optitrack system \cite{optitrack}.





\subsection{Participants and Procedure}
\label{sec:us1:proc}


The experiment was conducted at a North American post-secondary institution, where the large curved display is located.
Ethics approval was granted by the institution.
We invited 12 right-handed participants (6 males and 6 females) from a local university to participate. 
They were, on average, 23.25 years old ($SD = 6.24$).
Only two participants had prior experience interacting with large wall-sized displays. 
Each participant was compensated with a \$15 honorarium for their time.

Participants first completed a demographics questionnaire focusing on their prior experience using wall-sized displays.
The pointing task was then described.
The study was separated into six blocks, one per user position tested.
Each block began with 15 practice trials to help participants familiarize themselves with the study conditions.
Participants then completed 60 successive trials (six task difficulties - i.e., IDs -  repeated 10 times) before taking a two-minute break and moving to the next position.
Each block ended with participants completing a NASA-TLX questionnaire for the current position.

User positions were counterbalanced to reduce ordering effects.
Participants each completed 360 trials (3 \textit{Interaction Distances} $\times$ 2 \textit{Lateral Offsets} $\times$ 2 \textit{Widths} $\times$ 3 \textit{Amplitude} $\times$ 10 repetitions).
The study lasted approximately 60 minutes on average.

The \textbf{task} consisted of a 2D Fitts' Law selection task using a hand-held controller. 
Participants used the controller to select targets by positioning the cursor (2.5 cm diameter) within the boundary of each target. 
Selections were invoked by pressing the controller's trigger button.
Each trial began with participants selecting a white start circle (20 cm radius) rendered at a random location on the display. 
Upon selection, the start circle would disappear, and a red circular target would appear on the display with a specific \textit{Width} and \textit{Amplitude} based on the trial's ID.
After selecting the red circle, the trial would end, and the system would present a new start circle.  
Participants were instructed to select the targets as quickly and as accurately as possible.
Selections made outside the target boundary were flagged as errors.

\subsection{Measurements}

We recorded movement time and success rate for each trial, i.e., how long it took a participant to select the red target after selecting the white start circle and if they successfully selected the target.
Any selection outside the target's boundary was considered an error.
Additionally, we recorded participants' subjective ratings about the task for every tested position through the NASA-TLX \cite{nasatlx} questionnaire.


\subsection{Results}


We analyze movement time with repeated measures ANOVA on log-transformed data and post-hoc pairwise comparisons with Bonferroni corrections. 
We use Wilcoxon Signed-Rank test for the error rate data. 
For all tests, we report Greenhouse-Geisser corrected p-values and degrees of freedom.
All reported confidence intervals (CIs) are 95\%.

\begin{figure*}[h]
	\centering
	\includegraphics[width=1\columnwidth]{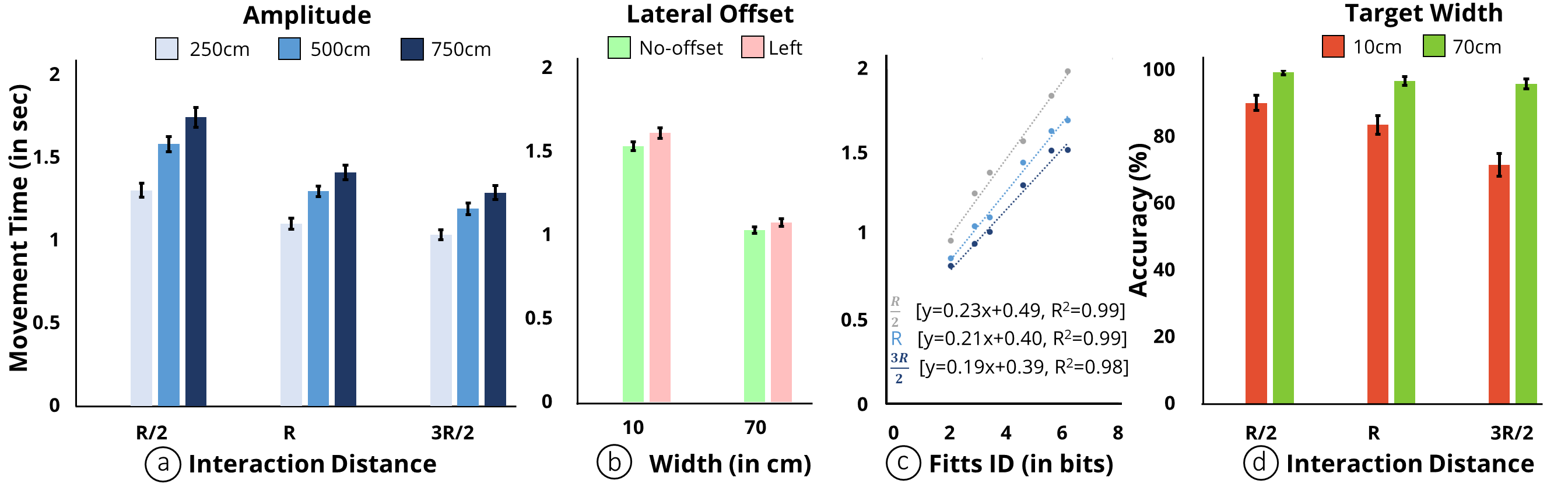}
	\caption{ 
 (a) Movement Time by \textit{Interaction Distance} for each target \textit{Amplitude}; 
 (b) Movement time by target \textit{Width} for each \textit{Lateral Offset}; (c) Fitts' Law regression lines for Movement Time by  \textbf{Fitts ID} for each \textit{Interaction Distance} and (d) Accuracy by \textit{Interaction Distance} for each target \textit{Width}.
 } 
	\label{fig:exp1-res} 
\end{figure*}

\subsubsection{Movement Time}


We found significant main effects of \textit{Amplitude} ($F_{2,22}=153.58$, $p < .0001$, $\eta^2$ = 0.93), \textit{Width} ($F_{1, 11}=250.92$, $p < .0001$, $\eta^2$ = 0.96), \textit{Interaction Distance} ($F_{2,22}=34.95$, $p< .0001$, $\eta^2$ = 0.76)  and \textit{Lateral Offset} ($F_{1,11}=6.92$, $p< .05$, $\eta^2$ = 0.39) on movement time (see Fig. \ref{fig:exp1-res}b-c). 
The mean movement time is 1.52s (CI: [1.49, 1.55]) for \textit{Interaction Distance} \(\frac{1}{2}\)R, 1.26s (CI: [1.23, 1.28]) for R and 1.16s (CI: [1.14, 1.18]) for \(\frac{3}{2}\)R. 
Post-hoc pairwise comparisons reveal that pointing at \(\frac{3}{2}\)R is significantly faster than \(\frac{1}{2}\)R and R (all $p<.0001$). 
In addition, pointing at R is significantly faster than \(\frac{1}{2}\)R ($p<.0001$). 
Furthermore, pointing from the No-offset (mean: 1.29, CI: [1.24, 1.34]) position is significantly faster than pointing from the Left-offset (mean: 1.36, CI: [1.30, 1.42]) position. 
Finally, we found a significant interaction effect of \textit{Amplitude} $\times$ \textit{Interaction Distance} ($F_{4,44}=11.45$, $p<.0001$, $\eta^2 = 0.51$) (see Fig. \ref{fig:exp1-res}c) and \textit{Width} $\times$ \textit{Interaction Distance} ($F_{2,22}=10.35$, $p<.001$, $\eta^2 = 0.51$). 


\subsubsection{Accuracy}

We found a significant main effect of \textit{Interaction Distance} on pointing accuracy ($\chi^2 (2)=20.17, p<.001$). 
Post-hoc pairwise comparisons revealed that pointing from \(\frac{1}{2}\)R (M = 0.95, [0.93, 0.96]) is significantly more accurate than R (M = 0.90, CI: [0.87, 0.93]) and \(\frac{3}{2}\)R (M = 0.84, CI: [0.8, 0.88]) (see Fig. \ref{fig:exp1-res}d). 
Furthermore, pointing from R is significantly more accurate than \(\frac{3}{2}\)R. 
We also found a significant main effect of \textit{Width} ($W=0, Z = -3.06, p<.001,r=-0.62$), with 0.1m-sized targets (M = 1.6, CI: [1.56, 1.64]) leading to less accurate results than 0.7m-sized ones (M = 1.04, CI: [1.01, 1.07]) (see Fig. \ref{fig:exp1-res}d).
Finally, we did not find a significant effect of \textit{Lateral Offset} ($p=0.62$) and \textit{Amplitude} ($p=0.56$) on accuracy.


\subsubsection{Fitts' Law Regression Lines}

We found that Fitts' Law can accurately model movement time for both \textit{Lateral Offsets} ($R^2\ge0.99$ for all). 
Interestingly, movement time predictions for \textit{Lateral Offsets}  have approximately the same slope, indicating that both no-offset ($y=0.21x+0.41$) and left-offset ($y=0.22x+0.42$) positions exhibit the same tolerance to an increase in task difficulty.
Additionally, we observed that Fitts' Law can accurately model movement time with varying \textit{Interaction Distances} ($all R^2\ge0.98$; See Fig. \ref{fig:exp1-res}a). 
Movement time is found to have a linear relationship with \textit{Interaction Distance} (see Fig. \ref{fig:exp1-res}a and \ref{fig:exp1-res}b), where larger distances result in faster selections when compared to shorter ones, e.g., \(\frac{3}{2}\)R leading to faster selections than R. 


\begin{figure}[htbp]
	\centering
	\includegraphics[width=1\columnwidth]{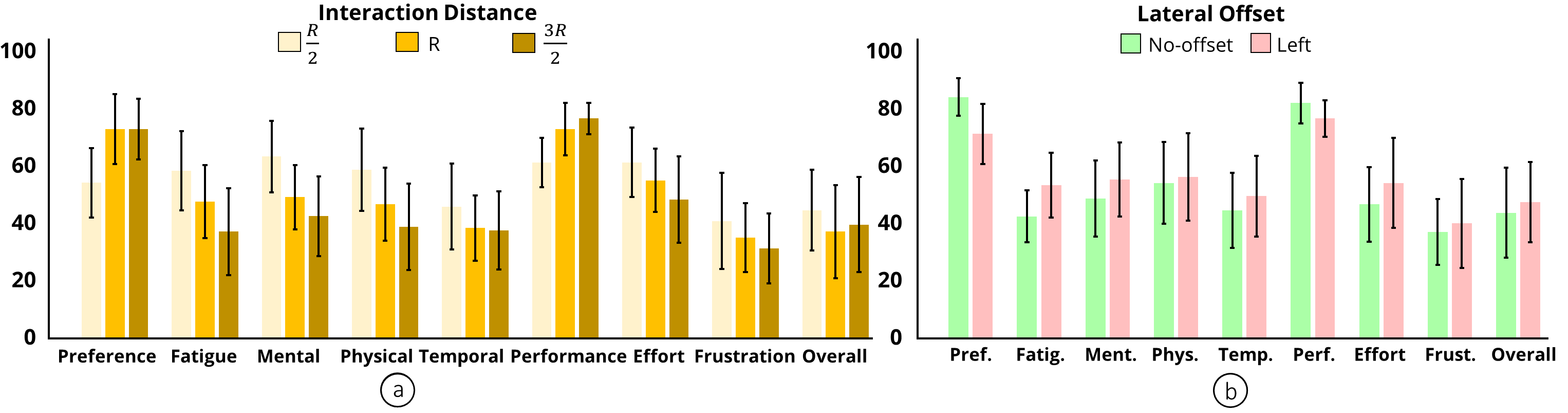}
	\caption{Subjective feedback score: (a) for each \textit{Interaction Distance}, and (b) for each \textit{Lateral Offset} (error bars represent are 95\% confidence intervals).
 }
	\label{fig:exp1-nasatlx} 
\end{figure}

\subsubsection{Preference Scores}

We collected users' feedback on the seven NASA-TLX \cite{nasatlx} criteria along with their preferences and perceived fatigue for each \textit{Interaction Distance} and \textit{Lateral Offset} condition. \update{We calculated the overall workload using raw NATA-TLX values. For enhanced readability, we inverted the performance scores such that a higher score indicates better performance.}


For \textit{Interaction Distances} \update{(see Figure \ref{fig:exp1-nasatlx}a)}, we found only significant differences for Preference ($\chi^2 (2,N=12)=8.21, p<.05$), Fatigue ($\chi^2 (2,N=12)=10.71, p<.01$), Mental Demand ($\chi^2 (2,N=12)=16.33, p<.001$), Physical Demand ($\chi^2 (2,N=12)=9.91, p<.01$), Performance ($\chi^2 (2,N=12)=6.44, p<.05$), and Frustration ($\chi^2 (1,N=12)=7, p<.05$). 
Pairwise comparisons revealed that \(\frac{3}{2}\)R has higher Preference scores than \(\frac{1}{2}\)R.
Additionally, \(\frac{1}{2}\)R has higher Fatigue, Physical Demand, and Mental Demand, as well as higher than both \(\frac{3}{2}\)R and R.

For \textit{Lateral Offsets} \update{(see Figure \ref{fig:exp1-nasatlx}b)}, we found participants significantly prefer being positioned in No-offset positions compared to Left-offset for three criteria: Preference ($\chi^2 (1,N=12)=5.33, p<.05$), Fatigue ($\chi^2 (1,N=12)=6.40, p<.05$), and Mental Demand ($\chi^2 (1,N=12)=4.50, p<.05$)]. 
We found no significant differences across all other dimensions.

\subsection{Discussion}

\subsubsection{Effects of Interaction Distance}

We found movement time to decrease linearly as the user moves farther away from the display (see Fig. \ref{fig:exp1-res}c), resulting in faster pointing speeds.
\textbf{We thus accept H1.}

We additionally found pointing accuracy to decrease as the user moved away from the display's centre (i.e., larger \textit{Interaction Distances}): 5\% error rate at \(\frac{1}{2}\)R, 10\% at R and 16\% at \(\frac{3}{2}\)R (see Figure \ref{fig:exp1-res}d). 
This result is expected as pointing at a distance introduces jitter from hand tremors \cite{kopper2011understanding}. 
The jitter is amplified with increases in \textit{Interaction Distances}, as the same angular jitter results in larger cursor displacements the further away the user, leading to larger error rates.
\textbf{We therefore accept H3.} 

The error rates based on \textit{Interaction Distance} further varied based on the target widths: smaller targets (20cm) at \(\frac{3}{2}\)R, R and \(\frac{1}{2}\)R have error rates of 28\%, 16\% and 10\%, while large targets (75cm) have error rates of 4\%, 3\% and 1\% respectively.

In summary, increased distances from the display correlates with quicker selection times but also higher error rates. 
We envision cursor enhancement techniques could reduce cursor speed at large \textit{Interaction Distances}, and increase it at short distances to provide a more uniform experience to users.


\subsubsection{Effects of Lateral Offset}

Contrary to prior work \cite{Ullah2023Pointing}, we observed participants were significantly faster at selecting targets from the no-offset positions compared to the left-offset positions. 
Furthermore, participants significantly preferred being positioned in No-offset positions compared to Left-offset based on Preference, Fatigue, and Mental Demand.
\textbf{We thus accept H2.}

\section{User Study 2 - Pointing Augmentations for Large Curved Displays} 

We confirmed in our first study the effects of \textit{Interaction Distance} on pointing performance on large curved displays, i.e., both movement time ($p < .0001$, with a large effect size) and accuracy ($p < .001$, with a large effect size).
Furthermore, we observed \textit{Lateral Offset} has a medium-sized effect on pointing speed ($p < .05$).
Ideally, users should be highly accurate and quick while pointing regardless of their position.
We thus see a need to design cursor-enhancement techniques for large curved displays to improve pointing performance from more difficult positions.
We thus conducted a within-subject user study to evaluate which of six motor- and visual-space-based enhancements best afford users with optimal pointing performance across positions around a large curved display. 



We define motor-space enhancements as techniques that modify cursor speed (i.e., Control Display (CD) gain) to facilitate moving towards a target, e.g. based on the controller's movement or user distance from the large display. 
On the other hand, visual-space enhancements modify the visual aspects of the cursor (e.g., its size) to facilitate target selection.
In this study, we consider three motor-space enhancement techniques:
1) "Pointer Acceleration" (PA) cursor, an acceleration-based technique \cite{nancel2013high}; 
2) "Position Based Acceleration" (PBA) cursor, where CD gain varies based on \textit{Interaction Distance}; 
and 3) "Pointer Acceleration and Distance-based" (PADIST), combining PA and PBA.
We further consider their visual-space-augmented variants, where the cursor size varies based on motor-space movement: 4) PASIZE; 5) PBASIZE; and 6) PADISTSIZE. 

Formally, this study consists of a 2D Fitts' Law selection task and involves three factors from the first study: \textit{Interaction Distances} (\(\frac{1}{2}\)R, R, and  \(\frac{1}{2}\)R), \textit{Lateral Offsets} (No-offset and Left-offset), as well as \textit{Target Amplitude} (250cm, and 750cm).
We further examine each of the six pointing enhancement techniques with respect to these factors.

We formulate the following Hypotheses regarding the pointing enhancement techniques:
\begin{itemize}
    \item \textbf{H1}) Pointing performance (movement time and accuracy) will be superior for visual-space-based enhancement techniques (i.e., PBASIZE, PASIZE, PADISTSIZE) compared to techniques without visual enhancements (i.e., PBA, PA, PADIST).
    \item \textbf{H2}) Techniques combining acceleration and \textit{Interaction Distance} will lead to superior pointing performance (movement time and accuracy) across user positions, compared to techniques that use only \textit{Interaction Distance} or acceleration.
\end{itemize}

\subsection{Techniques}

Numerous techniques have been proposed to enhance pointer speed and acceleration \cite{casiez2008impact, nancel2013high}. 
Such techniques typically involve modifying the Control Display (CD) gain or cursor speed, representing the ratio between cursor movement and the input-device movement. 
For our techniques, we choose Nancel et al.'s \cite{nancel2013high} equation to change CD gain and cursor size based on motor-space movements as the formula allows mapping an output value (e.g., cursor speed and cursor size) based on different motor-space changes (e.g., controller movement and \textit{Interaction Distance}).
In our study, we consider users' motor-space movement in the 3D space and the \textit{Interaction Distance} of the user as input to generate the CD gain and the current cursor size as outputs.
We evaluate pointing performance using six cursor enhancement techniques combining visual-space and motor-space enhancements (see Table \ref{tab:cursors}).
Each technique modifies the cursor speed based on the user's \textit{Interaction Distance}, or motor-space movement.
We focus exclusively on \textit{Interaction Distance} and do not consider \textit{Lateral Offset}, as we found the former to have much larger effects on performance and we wish to control the complexity of our study.
Visual-space-augmented variants further modify the cursor's size from a 2.5cm diameter to a 20cm diameter based on controller movement.


We define each motor-space enhancement using Nancel et al.'s \cite{nancel2013high} CD-gain equations, which result in cursor speeds increasing along with motor-space speed (e.g., slow motor-space movements will result in slower cursors affording more precise pointing): 

\begin{equation} \label{eq:CDgain-formula}
CD(x)=\frac{CD_{\max }-CD_{\min }}{1+e^{-\lambda\left(x -  V_{\mathrm{inf}}\right)}} +CD_{\min },
\end{equation}

where CD(x) is the augmented speed of the cursor based on the controller's speed $x$.
$CD_{\max}$ and $CD_{\min}$ correspond to the maximum and minimum speed the cursor can reach. 
Finally, $\lambda$ is a non-negative value defining the slope of the equation at its inflection point $V_{\text {inf }}$. 
In our study, we set the value of $\lambda$ to 20 based on the results of pilot studies. 
The value of $V_{\text {inf }}$ can be derived as follows: 

\begin{equation} \label{eq:find-v-inf-formula}
V_{\text {inf }}=\text { r }_{\text {inf }} \cdot\left(V_{\max }-V_{\min }\right)+V_{\min },
\end{equation}

where $V_{\max}$ and $V_{\min}$ correspond to the maximum and minimum controller speeds, and $ \text {r}_{\text {inf }}$ scales the value of $V_{\text {inf }}$.
All three variables are constants empirically chosen based on the pointing environment. 
In our case we used $V_{\max}$ = 1, $V_{\min}$ = 0.1 and $r_{inf}$ = 0.5 based on pilot studies.  

We further adapt Equations \ref{eq:CDgain-formula} and \ref{eq:find-v-inf-formula} to consider \textit{Interaction Distance} by modifying the definitions of $CD_{\max}$ and $CD_{\min}$:
\begin{equation} \label{eq:proposed}
\begin{split}
CD_{max} = \overline{CD}_{max}-(a * d_s), \\ 
CD_{min} = \overline{CD}_{min}-(a * d_s),
\end{split}
\end{equation}

where $d_s$ corresponds to the scaled \textit{Interaction Distance}, which we set to 0 at 0.5R, 0.5 at 1R, and 1 at 1.5R.
$a$, $\overline{CD}_{max}$, and $\overline{CD}_{min}$ are constants which values are empirically chosen to define an optimal range of CD-Gain.
We set their values to be equal to 0.2, 1.2 and 0.8 respectively based on the results of a pilot study.
\begin{table}[h]

 \begin{adjustbox}{width=0.75\textwidth}
\begin{tabular}{c|ccc}
  {\textbf{Visual-Space} }  & \multicolumn{3}{c}{\textbf{Motor-Space} } \\ 
                         & \multicolumn{1}{c}{Acceleration} & \multicolumn{1}{c}{Distance } & Acceleration $+$ Distance \\ \hline 
 \multicolumn{1}{c|}{None}   & \multicolumn{1}{c}{PA}  & \multicolumn{1}{c}{PBA}          & PADIST                          \\ 
\multicolumn{1}{c|}{Cursor Enlargement}      & \multicolumn{1}{c}{PASIZE}  & \multicolumn{1}{c}{PBASIZE}      & PADISTSIZE                      \\ 
\end{tabular}
\end{adjustbox}
\caption{Cursor-enhancement technique design space.}
\label{tab:cursors}
\end{table}


Finally, we note existing pointer enhancement techniques for flat displays must be adapted to extend to curved displays, due to different geometry.
Traditional methods for flat displays, which rely on two-dimensional transformations to navigate cursor movement, cannot be used as-is on curved displays because they do not consider the display's depth and curvature.
We thus adapt existing approaches to account for these factors.
Specifically, we track the user's input device in space and calculate the cursor movement as a function of the controller's rotations from the user's current position (see Appendix \ref{appendix} for more information).

\subsubsection{PA: Pointer Acceleration-based} 

The Pointer Acceleration-based (PA) technique \cite{nancel2013high} scales cursor speed based on controller speed, as defined by Equations \ref{eq:CDgain-formula} and \ref{eq:find-v-inf-formula} with $CD_{max}=1.2$ and $CD_{min}=0.8$.
This technique helps the user to move across large pointing distances faster, and facilitates smaller target selections by slowing down the cursor during slower motor-space movements.
However, this technique does not consider how pointing performance may vary based on user position. 

\subsubsection{PASIZE: Pointer Acceleration-based Cursor Enlargement} 

PASIZE augments the PA technique with cursor enlargement defined using Equation \ref{eq:CDgain-formula} with $CD_{max}=20$ and $CD_{min}=2.5$ corresponding to the maximum and minimum size of the cursor in centimeter. 
This results in the cursor's size increasing when performing faster motor-space movements to facilitate target selections.
We expect this technique to result in faster movement times than PA, as users will not have to rely as much on slow movements for selections. 
However, like PA, this technique does not account for changes in pointing performance based on user position. 

\subsubsection{PBA: Position based Acceleration}  

PBA modifies the cursor's speed purely based on the user's \textit{Interaction Distance}.
This change is calculated using Equations \ref{eq:CDgain-formula}, \ref{eq:find-v-inf-formula}, and \ref{eq:proposed} where we use the value of \textit{Interaction Distance} in meters as $x$ and set $CD_{max}=4.5$ meters and $CD_{min}=0.7$ meters. 
Given that shorter \textit{Interaction Distances} have led to more accurate selections in our first study, PBA increases cursor speed at short distances and slows it down at longer distances.
In adjusting the speed-accuracy trade-off of pointing, we expect pointing performance using PBA to be more balanced across user positions.
However, it may further result in more difficult selections at short distances, especially when targets are small. 

\subsubsection{PBASIZE: Position based Acceleration with Cursor Enlargement}
This technique augments PBA by changing the cursor size based on Equation \ref{eq:CDgain-formula}, using the user's motor-space movement.
Like PASIZE, cursor size is set to vary between 2.5 and 20cm.
We expect PBASIZE to further facilitate pointing compared to PBA, but to also suffer from increased pointing difficulty at shorter distances.

\subsubsection{PADIST: Pointer Acceleration and Distance-based} 
PADIST combines both augmentations of the PA and PBA techniques to modify cursor speed.
$CD_{max}$ and $CD_{min}$ are calculated using Equation \ref{eq:proposed} with the user's \textit{Interaction Distance.}
These values are then used in Equations \ref{eq:CDgain-formula} and \ref{eq:find-v-inf-formula} with the user's motor-space movement to obtain the cursor's speed.
We expect this method to leverage both PA and PBA's advantages to provide superior pointing performance while minimizing their disadvantages.
However, its use may be more frustrating if the parameter values of Equation \ref{eq:proposed} do not align with the environment, e.g., a small $CD_{min}$ value resulting in much slower selections. 

\subsubsection{PADISTSIZE: Pointer Acceleration and Distance-based Cursor Enlargement}

PADISTSIZE augments PADIST by altering cursor size based on the user's motor-space movement.
This change is calculated using \ref{eq:CDgain-formula} with $CD_{max}=20$ and $CD_{min}=2.5$ corresponding to the maximum and minimum size of the cursor in centimeter.
PADISTSIZE combines controller acceleration and \textit{Interaction Distance} along with visual-space enhancement (i.e., cursor enlargement).
In doing so, we expect PADISTSIZE to outperform other techniques due to its richer set of augmentations.
However, we also acknowledge that its performance is highly dependent on the right values used in Equation \ref{eq:proposed}. 


\subsection{Apparatus and Implementation}

We used the same large curved display and controller setup as in Study 1.
The Unity application was modified to accommodate the six techniques for pointing at the targets across different user positions.

Selection targets had a width of 10cm, and were distant by 250cm or 750cm (\textit{Target Amplitudes}).
This resulted in two distinct Index of Difficulty for trials: 4.70 and 6.25 bits.
We chose these values based on pilot studies to make the pointing task difficult enough for participants to make selection errors.
The number of IDs was reduced to keep the average session duration under 90 minutes.

\subsection{Participants and Procedure}

We invited 12 right-handed participants (7 males, 5 females) from a local university to participate. 
They were on average 24.16 years old ($SD = 4.65$ years).
Few participants had prior experience (M = 0.25 years, SD = 0.32 years) interacting with large displays, and none had experience interacting specifically with large curved displays.

The procedure follows the same steps and task as Study 1 (see Section \ref{sec:us1:proc}). 
The experiment was conducted at a North American post-secondary institution, where the large curved display is located.
Ethics approval was granted by the institution.

Participants completed the pointing task at each position for a specific technique, before moving to a new technique.
NASA-TLX questionnaires were completed for each technique only.
Techniques were counter-balanced to reduce ordering effects.
Participants each completed 720 trials (6 \textit{Techniques} $\times$ 3 \textit{Interaction Distances} $\times$ 2 \textit{Lateral Offsets} $\times$ 2 \textit{IDs} $\times$ 10 repetitions).
The study lasted approximately 90 minutes on average.
Each participant was compensated with a \$15 honorarium for their time.

We recorded demographic information, movement time, success rate, and subjective feedback ratings for each of the techniques.

\subsection{Results}


We analyze movement time with repeated measures ANOVA on log-transformed data and post-hoc pairwise comparisons with Bonferroni corrections. 
We use the Wilcoxon Signed-Rank test to analyze accuracy. 
For all tests, we report Greenhouse-Geisser corrected p-values and degrees of freedom.
All reported confidence intervals (CIs) are 95\%.

\begin{figure}[htbp]
	\centering
	\includegraphics[width=1.0\columnwidth]{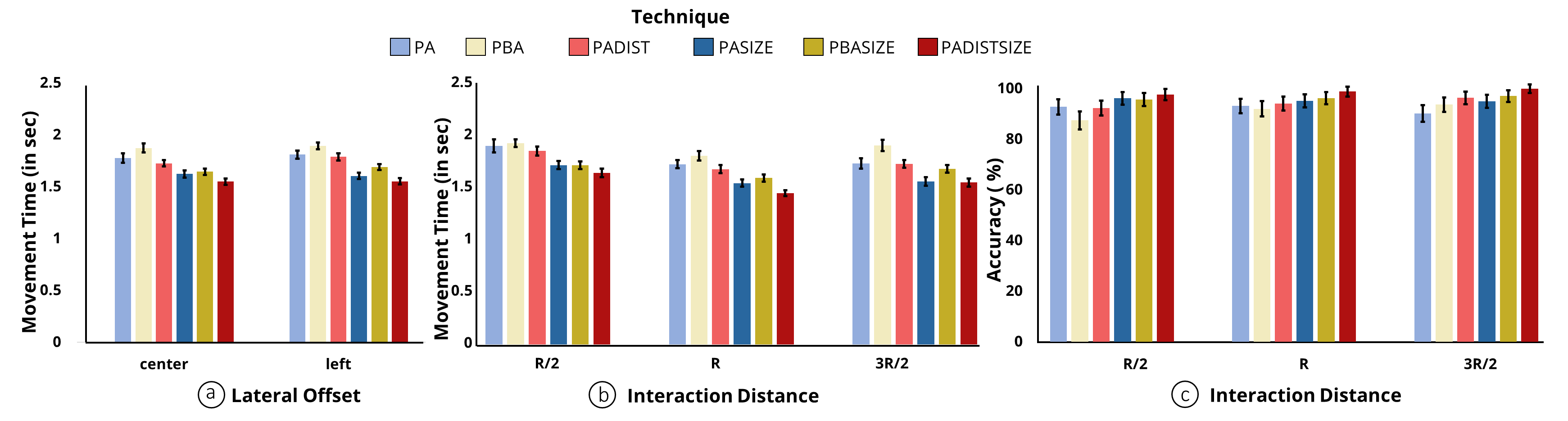}
	\caption{ For each \textbf{Technique}, we show average (a) Movement time by \textit{Lateral Offset}, (b) Movement time by \textit{Interaction Distance} and (c) Accuracy by \textit{Interaction Distance}. Error bars represent  95\% confidence intervals.}

	\label{fig:exp2-res} 
\end{figure}

\subsubsection{Movement Time}

We found significant main effects of \textit{Technique} ($F_{5,55}=13.36$, $p < .001$, $\eta^2$ = 0.55), \textit{Interaction Distance} ($F_{2,22}=44.47$, $p < .001$, $\eta^2$ = 0.80), \textit{Amplitude} ($F_{1,11}=365.92$, $p < .001$, $\eta^2$ = 0.97), and \textit{Lateral Offset} ($F_{1,11}=5.97$, $p<.05$, $\eta^2$ = 0.35) (see Fig. \ref{fig:exp2-res}a) on movement time. 

For the \textit{Techniques} (see Fig. \ref{fig:exp2-res}a), the mean movement time is 1.77s (CI: [1.72, 1.82]) for PA, 1.87s (CI: [1.81, 1.92]) for PBA, 1.75s (CI: [1.70, 1.80]) for PADIST, 1.60s (CI: [1.55, 1.65]) for PASIZE, 1.66s (CI: [1.62, 1.71]) for PBASIZE, and 1.54s (CI: [1.50, 1.58]) for PADISTSIZE. 
Post-hoc pairwise comparisons reveal that the  \textit{Techniques} with visual-space enhancements are significantly faster than their non-visually-enhanced counterparts, i.e., PADISTSIZE, PASIZE and PBASIZE have faster movement times than PADIST, PA and PBA (all $p<.001$). 
We further find the PBA techniques to be the slowest out of the three types of motor-space enhancements (all $p<.001$).
For the visual-space enhancements, PADISTSIZE is significantly faster than PBASIZE ($p<.0001$) and PASIZE ($p<.05$).

Regarding the effect of \textit{Interaction Distance} (see Fig. \ref{fig:exp2-res}b), we find the mean movement times to be 1.79s (CI: [1.75, 1.83]) for \(\frac{1}{2}\)R, 1.63s (CI: [1.60, 1.67]) for R, and 1.68s (CI: [1.64, 1.71]) for \(\frac{3}{2}\)R.
Post-hoc pairwise comparisons reveal that pointing from R is significantly faster than \(\frac{1}{2}\)R and \(\frac{3}{2}\)R (all $p<.0001$). 
In addition, pointing at \(\frac{3}{2}\)R  is significantly faster than \(\frac{1}{2}\)R ($p<.0001$). 
In terms of the \textit{Lateral Offset}(see Fig. \ref{fig:exp2-res}a), the no-offset position (mean time 1.68s, CI: [1.65, 1.72]) was significantly faster ($p<.0001$) than the Left-offset position (mean time 1.71s, CI: [1.68, 1.74]). 

Finally, we found several interaction effects: \textit{Amplitude} $\times$ \textit{Interaction Distance} ($F_{2,22}=9.34$, $p<.01$, $\eta^2 = 0.46$) and \textit{Amplitude} $\times$ \textit{Lateral Offset} ($F_{1,11}=13.66$, $p<.01$, $\eta^2 = 0.55$).
We observed similar trends for both high and low amplitudes in terms of \textit{Interaction Distance} (all $p<0.05$): \(\frac{3}{2}\)R is the fastest followed by R and then followed by \(\frac{1}{2}\)R. 
For the low amplitudes (250cm), the no-offset position was faster in terms of selection time compared to the Left-offset position( $p<.0001$). 
However, for the higher amplitude (A=750cm), we did not find a significant effect of \textit{Lateral Offset} on Movement time ($p=.11$).  




\subsubsection{Accuracy}

We found a significant effect of \textit{Technique} ($\chi^2 (5)=23.1, p<.001$) on pointing accuracy (see Fig. \ref{fig:exp2-res}c).
The mean accuracy rate is 0.89 (CI: [0.86, 0.92]) for PA,  0.88 (CI: [0.85, 0.91]) for PBA, 0.91 (CI: [0.88, 0.94]) for PADIST, 0.92 (CI: [0.89, 0.96]) for PASIZE, 0.93 (CI: [0.90, 0.96]) for PBASIZE, and 0.95 (CI: [0.94, 0.97]) for PADISTSIZE. 
Pairwise comparisons revealed that only PADISTSIZE is significantly more accurate than PA ($p<.05$, $r = .68$)and PBA ($p<.05$, $r = .71$).
We also found a significant effect of\textit{Amplitude} between 250cm (M=0.93, CI: [0.91, 0.94]) and 750cm (M=0.90, [0.88, 0.93]) on accuracy ($W = 77; Z = 2.98; p<.001, r = .61$). 
However, we did not observe a significant effect of \textit{Interaction Distance} ($p=.08$) or \textit{Lateral Offset} ($p=.20$) on accuracy.


\subsubsection{Fitts' Law Throughput Analysis}

We evaluated the pointing performance of each technique based on throughput \cite{fitts1954information, iso20009241, zhai2004speed, mackenzie1992fitts}. 
We used Crossman’s correction \cite{crossman1957speed} to mitigate potential biases \cite{zhai2004speed}, and used the means-of-means methodology \cite{soukoreff2004towards, iso20009241} to calculate throughput for each technique \cite{wobbrock2011effects}. 

We found significant main effects of \textit{Technique} ($F_{5,55}=4.11$, $p < .01$, $\eta^2$ = 0.27) on throughput (see Fig. \ref{fig:exp2-nasatlx}b).
We also found significant main effects of \textit{Interaction Distance} ($F_{2,22}=35.65$, $p < .001$, $\eta^2$ = 0.76), \textit{Amplitude} ($F_{1,11}=8.51$, $p < .05$, $\eta^2$ = 0.44), and \textit{Lateral Offset} ($F_{1,11}=5.97$, $p<.05$, $\eta^2$ = 0.35) on throughput. 
We found that the mean throughput ($TP$) is 2.97 bits/s (CI: [2.77, 3.18]) for PA, 2.76 bits/s (CI: [2.71, 3.04]) for PBA, 3.01 bits/s (CI: [2.80, 3.22]) for PADIST, 3.01 bits/s (CI: [2.79, 3.23]) for PASIZE, 2.87 bits/s (CI: [2.71, 3.04]) for PBASIZE, 3.11 bits/s (CI: [2.90, 3.31]) for PADISTSIZE. 
Pairwise comparisons reveal that PA ($p<.05$), PADIST, ($p<.01$), PASIZE ($p<.05$) and PADISTSIZE ($p<.001$) have significantly higher throughput than PBA.
Additionally, PADIST ($p<.05$) and PADISTSIZE ($p<.01$) have significantly higher throughput than PBASIZE.
For \textit{Lateral Offset} ($F_{1,11}=2.19$, $p=0.17$), we did not observe any significant effect on movement time.
We also did not find any interaction effects between the studied factors.




\subsubsection{Preference Scores}

\begin{figure}[h]
	\centering
	\includegraphics[width=1\columnwidth]{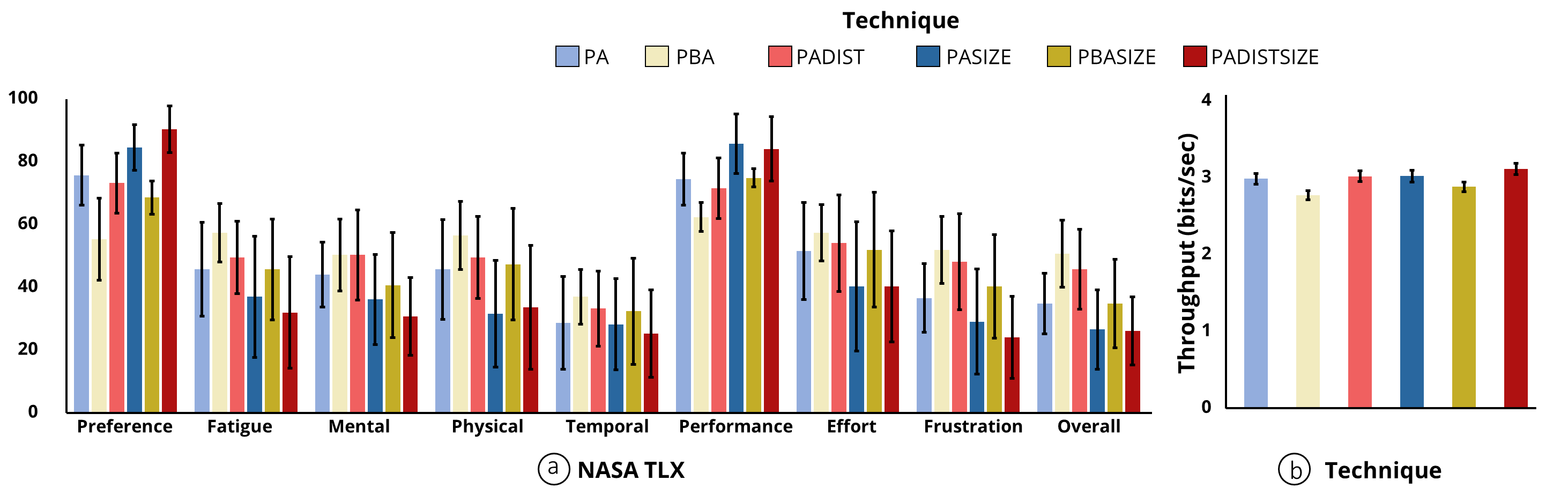}
	\caption{(a)Subjective feedback score (NASA-TLX \cite{nasatlx}) for each \textit{Technique} and (b) For each \textbf{Technique}, we show average Throughput.
 Error bars represent 95\% confidence intervals.} 
	\label{fig:exp2-nasatlx} 
\end{figure}

We collected participant feedback on the seven NASA-TLX criteria along with their preferences across all \textit{Techniques} (see Fig. \ref{fig:exp2-nasatlx}a). \update{We calculated the overall workload using raw NASA-TLX values, but for clarity, we inverted the performance scores.}
We found significant differences for all metrics: Preference ($\chi^2 (5,N=12)=34.88, p<.001$), Fatigue ($\chi^2 (5,N=12)=22.80, p<.001$), Mental Demand ($\chi^2 (5,N=12)=24.97, p<.001$), Physical Demand ($\chi^2 (5,N=12)=33.22, p<0.001$) , Temporal Demand ($\chi^2 (5,N=12)=20.21, p<.001$), Performance ($\chi^2 (5,N=12)=34.76, p<.001$), Effort ($\chi^2 (5,N=12)=18.73, p<.01$), Frustration ($\chi^2 (5,N=12)=37.38, p<.001$), and Overall task load ($\chi^2 (5,N=12)=34.96, p<0.001$).
Pairwise comparisons revealed PADISTSIZE and PASIZE have higher Preference and Performance scores and lower Frustration scores than PBA.
Furthermore, PASIZE and PADISTSIZE were rated having lower mental demand, physical demand, and overall workload than PBA and PADIST.
Finally, PADISTSIZE had lower fatigue and effort ratings than PBA. 

\subsection{Discussion}

\subsubsection{Benefits of Visual-space Augmentations}

We found visually-augmented techniques to lead to faster, and more accurate pointing (by $4\%$), compared to their strictly motor-space enhancement counterparts.
The main advantage of visual-space enhancements is of facilitating selections by enlarging the cursor or targets and requiring less precise movements by the user. 
This suggests visual-space enhancements, like Bubble Cursor \cite{GrossmanCHI2005} may be promising directions to support pointing on large  curved displays.
We can however not suggest a specific enhancement, as we only tested one such technique in this study.

\subsubsection{Benefits of Combining Visual- and Motor-space Augmentations}

Although previous studies have solely focused on either visual-space enhancement \cite{GrossmanCHI2005} or motor-space enhancement \cite{nancel2013high}, our results suggest that for large curved displays, combining both visual and motor-space enhancement (i.e., PADISTSIZE) leads to superior performance in terms of movement time, accuracy and throughput.
For instance, we found that PADISTSIZE ($M = 95\%$) led to the most accurate selections on average, and is significantly more accurate than PBA ($M = 88\%$) and PA ($M= 89\%$).
However, we note such techniques may require more advanced parameter tuning to work well across contexts.

Considering all six techniques, we find PADIST and PADISTSIZE to outperform the others.
There are thus clear benefits to account for users \textit{Interaction Distance} in addition to acceleration-based augmentations.
For example, analyzing strictly motor-space enhancements, we found PADIST and PA  to result in $6\%$ and $5\%$ faster selections than PBA ($M = 1.87s$).
Similarly, comparing only the visual-space augmented techniques, selections using PADISTSIZE to be $4\%$ faster than PASIZE, and $7\%$ faster than PBASIZE.
Even in terms of throughput, which penalizes off-center selections more commonly encountered with visual-space augmented techniques, we find PADISTSIZE to achieve the higher throughput.

In conclusion, the use of acceleration and distance-based augmentations combined with visual-space augmentations (i.e., PADISTSIZE and PADIST) show promising potential for enhancing user interaction on large curved displays. 
Their effectiveness makes them suitable candidates for future studies or practical implementations in real-world scenarios.

\subsubsection{Effects of User Position}

We find the effect of \textit{Interaction Distance} on movement time to be different than in the first study: 
here, movement times were nearly the same at R distance and \(\frac{3}{2}\)R. 
Separating the times per technique, pointing had faster selection times at R \textit{Interaction Distance} compared to at \(\frac{3}{2}\)R for techniques which adjusted CD gain values based on \textit{Interaction Distance} (i.e., PBA, PBASIZE, PADIST, PADISTSIZE). 
This shows the utility of \textit{Interaction Distance}-based augmentations in affording faster selections closer to the display.
However, we note further development is needed to support interactions at shorter distances (e.g., \(\frac{1}{2}\)R).

In terms of \textit{Lateral Offset}, similar to Study 1, we found pointing from the no-offset position was significantly faster than from the left-offset position. 
Faster selections from the centered positions were also observed for all cursor enhancement techniques except PADISTSIZE and PASIZE (see Fig. \ref{fig:exp2-res}a).
This suggests that changes in CD gain can notably modulate the effects of position-based factors on pointing speed and accuracy.

\section{Design Implications, Limitations and Future Work}

\subsection{Design Implications}

\subsubsection{Effects of Interaction Distance}

We observed significant differences in pointing performance on large curved displays based on the user's \textit{Interaction Distance} in our first study: the farther away the user from the display, the faster and less accurate their pointing becomes.
For example, pointing from \textit{Interaction Distance} \(\frac{3}{2}\)R compared to \(\frac{1}{2}\)R leads to 16\% increase in speed but  11\% decrease in accuracy. 
However, in the second study, we observed faster selection times for \textit{Interaction Distances} of R compared to \(\frac{3}{2}R\), and no significant differences on accuracy for our \textit{Interaction Distance}-based enhancement techniques. 
Our proposed techniques thus afforded accurate and fast pointing across different user positions. 
We note  our evaluation focused on three specific distances, and may not generalize to much smaller or larger \textit{Interaction Distances} (e.g., 3R).
For instance, the implementation of our cursor enhancement techniques may require different parameter values. 
We argue researchers should test their implemented techniques from diverse \textit{Interaction Distances} and \textit{Lateral Offsets} to provide optimal interaction across different locations.

\subsubsection{Effects of Lateral Offset}

Across both studies, pointing from a centered position was consistently faster than from an off-centered position. 
We however did not observe any differences in terms of pointing accuracy.
Our results thus suggest benefits to having users interact with large curved displays when positioned along its main axis, but further research is necessary to assess the effects of \textit{Lateral Offsets} in more realistic settings.
Additionally, future research should investigate if specific conditions, postures, or user demographics might be more sensitive to lateral displacements, to ensure a more representative and efficient design approach.

\subsubsection{Motor-Space and Visual-Space Enhancements}

Our second study informs on the impacts of different enhancement techniques on pointing on large curved displays. %
All techniques led to faster movement times and higher accuracy compared to absolute pointing used in Study 1. 
Amongst the three motor enhancement techniques (PA, PBA and PADIST), acceleration-based techniques (i.e., PA and PADIST) affected pointing movement time.
Combining acceleration and \textit{Interaction Distance} (PADIST) yielded notably the best-pointing performance in terms of throughput.
We further note the benefits of considering \textit{Interaction Distance} in enhancement techniques, as such techniques minimized the movement time differences between the two larger distances.
We therefore suggest integrating both acceleration and \textit{Interaction Distance} in control schemes, as this combination notably demonstrates superior performance compared to methods relying on a single feature, such as PA and PBA.

Additionally, we observed that augmenting motor-space enhancements with a visual-space enhancement resulted in improved pointing performance across user positions.
Hence, we recommend combining visual-space and motor-space enhancements for optimal pointing performance on large curved displays.

\subsection{Limitations and Future Work}

Our two studies investigated a limited range of six user positions - three \textit{Interaction Distances} and two \textit{Lateral Offsets}.
\update{While this provided valuable insights, the inherent complexities introduced by the curved nature of the display and the dynamic positioning of the user were not fully captured in this study, potentially limiting the generalizability of our findings. Specifically, as users move closer to or further from the display or shift laterally, the angular width of targets changes due to varying interaction distances. Additionally, the actual distance from the user to different points on the screen varies non-linearly along the curvature (e.g., center vs. extreme sides), complicating interactions further. In this study, we employed linear measures for target width, target amplitude, and interaction distance. Future research should, therefore, focus on exploring these factors in terms of angular dimensions and non-linear distances that more accurately reflect the curved display context.} 

We acknowledge the range of user positions in real conditions are much richer than our tested conditions.
Future studies may investigate a broader range of user positions to derive a more comprehensive understanding of how spatial configurations influence user interaction, particularly in terms of pointing performance.
Furthermore, we conducted our studies with strictly right-handed participants to mitigate any potential confounding effects of handedness. 
However, this has constrained our understanding of how handedness may influence pointing on largue curved displays.
Future studies could incorporate participants of both right and left-handed orientations and explore both left and right \textit{Lateral Offsets}.
We also note we only considered hand-held controllers as the input modality, so our findings may not generalize as-is to other modalities for ray-casting such as head rotations, and eye gaze.

Moreover, our investigation focused only on pointing at 2D targets. 
Consequently, our insights may not generalize to other interactions on large curved displays, such as visual searches. 
Investigating other forms of interactions will provide a more comprehensive understanding of user experience on large curved displays.
Furthermore, it will be beneficial to examine how our findings extend to real-world applications, such as gaming, visual analytics, or entertainment, as these settings could present unique challenges and considerations not accounted for in isolated tasks. 
Lastly, our exploration was conducted on only one large semi-circular curved display with a specific curvature radius (i.e., 3270R and 3m tall) in a private room.
Our findings may thus not generalize to other curvature radii and types of curved displays.

\section{Conclusion}
In this paper, we focused on investigating how the position of a user affects their experience interacting with large curved displays and explored motor- and visual-space-based enhancement techniques to improve pointing performance across different positions.
Our first study revealed a notable trade-off between pointing speed and accuracy relative to the user's position from the display. 
The results showed that the farther away users stand, the faster they point, but with less accuracy. 
Faster pointing was also observed from no-offset positions compared to left-offset positions. 
Additionally, we discovered that adapting pointing interactions to the users' motor-space speed andthe distance from the display, combined with cursor enlargement techniques significantly improve pointing performance on large curved displays.
Finally, we provided recommendations to improve pointing performance on large curved displays under varied user positions. 
Taken together, our work contributes to our understanding on the design space of large curved displays for pointing interactions, and offers guidelines to design cursor enhancement techniques in this setting. 

\phantomsection
\addcontentsline{toc}{section}{Acknowledgements}
\section*{Acknowledgements}
This research was partially funded by NSERC Discovery Grant \#RGPIN-2019-05211 and the Canada Foundation for Innovation (CFI) Project number \#40440.

\appendix
\section{C\# Unity Code Snippet for Cursor Speed Enhancement in 3D Space via Game Object Rotation} \label{appendix} When using the Pointer Acceleration-based (PA) technique, the method ``PA()'' is called every frame inside the Update() function. The method for other techniques was similar, but the calculation of ``PAgetSpeed()'' differed based on each technique.
The current game object (in the code, ``transform.rotation'') was rotated in 3D space based on the movement of the controller (in the code, ``CGS.ControllerPosition.transform''). A cursor was placed in the orthogonal direction of the current game object. We rotate the current game object more for a faster speed than the controller, and rotate the current game object slower if we want to reduce the speed compared to the controller’s movement. This speed is determined by the function ``PAgetSpeed()''.

\begin{lstlisting}[
    caption={C\# Unity Method to Rotate the Current Game Object Based on the Controller's Rotation in Front of a Curved Display}, 
    label={lst:rotation-code},
    xleftmargin=1.5em,  % Adjust this value to increase the left margin
    xrightmargin=1.5em  % Adjust this value to increase the right margin
]
void PA()
{   
    Quaternion rotationDifference = CGS.ContollerPosition.transform.rotation * Quaternion.Inverse(lastcontrollerrotation);
    // Extract the angle and axis from the rotation difference.
    rotationDifference.ToAngleAxis(out float angle, out Vector3 axis);
    // current gameobject's rotation = rotation difference * speed factor.
    Quaternion tempRotation = transform.rotation * Quaternion.AngleAxis(angle * PAgetSpeed(), transform.InverseTransformDirection(axis));
    // Convert the rotation to Euler angles.
    Vector3 euler = tempRotation.eulerAngles;
    // Unity uses degrees from 0 to 360. Convert them to -180 to 180.
    if (euler.x > 180)
        euler.x -= 360;
    if (euler.y > 180)
        euler.y -= 360;
    // based on our display xMax = 90, xMin = -90, yMax = 70, and yMin = -70
    if ((euler.y <= xMax && euler.y >= xMin) && (euler.x <= yMax && euler.x >= yMin))
    {
        transform.rotation = tempRotation; // cursor always in display
    }
    // Store the controller's current rotation for the next frame.
    lastcontrollerrotation = CGS.ContollerPosition.transform.rotation;
}
\end{lstlisting}

\begin{lstlisting}[
    caption={Movement Speed of the Game Object rotation method is determined by the PAgetSpeed() Method}, 
    label={lst:rotation-code}, 
    xleftmargin=1.5em,  % Adjust this value to increase the left margin
    xrightmargin=1.5em  % Adjust this value to increase the right margin
]
float PAgetSpeed()
{
    float speed = 0;
    Vector3 mouseDelta = CGS.ContollerPosition.transform.position;
    Quaternion currentControllerRotation = CGS.ContollerPosition.transform.rotation;
    // Calculate the difference in position between the current and last frame
    Vector3 diff = mouseDelta - lastMousePosition;
    // Calculate the speed of the controller's movement (magnitude of diff / time since last frame)
    float controllerSpeed = diff.magnitude / Time.deltaTime;
    
    // The values below are constant values which are set based on pilot studies
    float x = controllerSpeed;
    float vInf = vRatioInf * (vMax - vMin) + vMin;
    speed = ((cdMax - cdMin) / (1 + Mathf.Exp(-lambda * (x - vInf)))) + cdMin;
    CGS.debug = "Controller speed: " + controllerSpeed + "\n" + "Cursor Speed: " + speed;
    lastMousePosition = mouseDelta;
    return speed;
}
\end{lstlisting}

\bibliographystyle{ACM-Reference-Format}
\bibliography{main.bib}

\end{document}